\documentclass[useAMS,usenatbib]{mn2e}
\usepackage{footnote,graphicx,color,multirow,amsmath,url,amssymb,tabularx,amssymb}
\usepackage{hyperref}

\hypersetup{
    colorlinks,
    citecolor=blue,
    filecolor=black,
    linkcolor=black,
    urlcolor=black
}

\def\lesssim{\mathrel{\hbox{\rlap{\hbox{\lower3pt\hbox{$\sim$}}}\hbox{\raise2pt\hbox{$<$}}}}}

\definecolor{refcol}{rgb}{0,0,1}
\def\refcol		{\color{refcol}}
\definecolor{minorcol}{rgb}{0,0,0}
\def\minor		{\color{minorcol}}

\begin{document}

\title[Quenching Histories of Fast and Slow Rotators]{SDSS-IV MaNGA: The Different Quenching Histories of Fast and Slow Rotators}
\author[Smethurst et al. 2017]{R. ~J. ~Smethurst,$^{1}$ K.~L.~Masters,$^{2}$  C. ~J. ~Lintott,$^{3}$ A.~Weijmans,$^{4}$  M.~Merrifield,$^{1}$ \newauthor S.~J.~Penny,$^{2}$ A. Arag\'on-Salamanca,$^{1}$  J.~Brownstein,$^{5}$ K.~Bundy,$^{6}$  N.~Drory,$^{7}$ \newauthor  D.~R.~Law,$^{8}$ R. ~C. ~Nichol $^{2}$ 
\\ $^1$ School of Physics and Astronomy, The University of Nottingham, University Park, Nottingham, NG7 2RD, UK
\\ $^2$ Institute of Cosmology and Gravitation, University of Portsmouth, Dennis Sciama Building, Barnaby Road, Portsmouth, PO13FX, UK 
\\ $^3$ Oxford Astrophysics, Department of Physics, University of Oxford, Denys Wilkinson Building, Keble Road, Oxford, OX13RH, UK
\\ $^4$ School of Physics and Astrononomy, University of St Andrews, North Haugh, St Andrews, Fife, KY169RJ, UK
\\ $^5$ Department of Physics and Astronomy, University of Utah, 115 S. 1400 E., Salt Lake City, UT 84112, USA
\\ $^6$ 	University of California, Santa Cruz, 1156 High St. Santa Cruz, CA 95064, USA
\\ $^7$ McDonald Observatory, The University of Texas at Austin, 1 University Station, Austin, TX 78712, USA
\\ $^8$ Space Telescope Science Institute, 3700 San Martin Drive, Baltimore, MD 21218, USA
\\
\\Accepted 2017 September 25. Received 2017 September 25; in original form 2017 August 25.
}

\maketitle

\begin{abstract}
Do the theorised different formation mechanisms of fast and slow rotators produce an observable difference in their star formation histories? To study this we identify quenching slow rotators in the MaNGA sample by selecting those which lie below the star forming sequence and identify a sample of quenching fast rotators which were matched in stellar mass. This results in a total sample of $194$ kinematically classified galaxies, which is agnostic to visual morphology. We use $u-r$ and $NUV-u$ colours from SDSS and GALEX and an existing inference package, \textsc{starpy}, to conduct a first look at the onset time and exponentially declining rate of quenching of these galaxies. An Anderson-Darling test on the distribution of the inferred quenching rates across the two kinematic populations reveals they are statistically distinguishable ($3.2\sigma$). We find that fast rotators quench at a much wider range of rates than slow rotators, consistent with a wide variety of physical processes such as secular evolution, minor mergers, gas accretion and environmentally driven mechanisms. Quenching is more likely to occur at rapid rates ($\tau \lesssim 1~\rm{Gyr}$) for slow rotators, in agreement with theories suggesting slow rotators are formed in dynamically fast processes, such as major mergers. Interestingly, we also find that a subset of the fast rotators quench at these same rapid rates as the bulk of the slow rotator sample. We therefore discuss how the total gas mass of a merger, rather than the merger mass ratio, may decide a galaxy's ultimate kinematic fate. 
\end{abstract}


\section{Introduction}\label{sec:intro}

Recent work studying the early-type (i.e. elliptical and lenticular) galaxy population has revealed that it is actually composed of two kinematically distinct populations. The majority of early-types are rotationally supported \citep{emsellem11} with $\sim7$ times the number of galaxies with kinematic discs (`fast' rotators), than those with either dispersion dominated kinematics (`slow' rotators) or kinematically decoupled cores \citep[which, along with slow rotators are collectively referred to as `non-regular' rotators;][]{cappellari07, emsellem07}.  This has led to the proposal of a revision of Hubble's morphological classification scheme in the form of a `comb' \citep[see Figure 24 of][]{cappellari16}, whereby the evolution of a galaxy, from  disc to bulge-dominated, takes place along a `tine' of the comb as a fast rotator, always retaining an underlying disc. If the discs of these regular rotators are destroyed, they then evolve along the `handle' of the comb to become slow rotators. 

\begin{table*}
\centering
\caption{Summary of the generalised rates of theorised internal and external quenching mechanisms (see \protect\citealt{smethurst17}).}
\label{table:qm}
\newcolumntype{C}{>{\centering\arraybackslash}X}%
\begin{tabularx}{\textwidth}{r|C|C}
\hline
\multicolumn{1}{l|}{}  & Internal Processes (`Nature')      & External Processes (`Nurture')     \\ \hline
Fast quenching         & AGN feedback            & Mergers           \\
Intermediate quenching & Mass quenching          & Environmental quenching \\
Slow quenching         & Morphological quenching & Gas accretion           \\ \hline
\end{tabularx}
\end{table*}

Dry major mergers are considered the most likely process to produce high stellar mass slow rotators~\citep{bois10, duc11, naab14} as they can rapidly destroy the disc dominated nature of a galaxy \citep{toomre72}. Low stellar mass slow rotators (i.e. dwarf ellipticals with $M_*~\lesssim~10^9~M_{\odot}$) are thought to be formed via harassment mechanisms in the group and cluster environment \citep{toloba15}. 
Fast rotators are thought to evolve from the slow build up of a galaxy's bulge over time, eventually overwhelming the disc. This growth is thought to occur via gas-rich major or minor mergers \citep{duc11} and by gas accretion \citep{cappellari13, johnston14} which can produce a bulge dominated but rotationally supported galaxy (which would be visually classified as an early-type in the Hubble classification scheme). 

The possible formation mechanisms listed above are also often proposed as external quenching mechanisms of star formation in a galaxy. However, these mechanisms are not thought to quench a galaxy at the same rate. Dynamically faster processes, such as mergers, are thought to quench star formation at rapid rates \citep{hopkins08a, snyder11, hayward14}, with major mergers thought to cause a much faster quench of the remnant galaxy than a minor merger \citep{lotz08b, lotz11}. Similarly, environmental processes, such as harassment, are also thought to cause quenching through repeated high speed interactions with neighbouring galaxies. Over time these interactions can strip both stars and gas from a galaxy and heat the gas needed for star formation \citep{knebe06, aguerri09}, quenching the galaxy at a slower rate than a merger. Slow quenching by an external process is also possible through gas accretion due to the large gravitational potential of the bulge which builds as the accreted gas sinks to the centre of the galaxy. This prevents the disc from collapsing and forming stars in an internal process which is categorised as morphological quenching \citep{martig09, fang13}. Similarly, there are internal processes which are theorised to cause quenching in galaxies, including AGN feedback \citep{croton06, somerville08}, mass quenching \citep{peng10, peng12} and morphological quenching \citep[e.g. due to a galactic bar,][]{zurita04, sheth05} at rapid, intermediate and slow quenching rates respectively. Crucially, external quenching processes are the only mechanisms theorised to be able to change the morphology of a galaxy \citep[see Section 1 of][for a more detailed introduction to possible quenching mehanisms]{smethurst17}. These quenching mechanisms and their theorised rates are summarised in Table~\ref{table:qm}.

If fast and slow rotators form via different mechanisms, we should therefore also expect to find a difference in the star formation histories of quenching or quenched fast and slow rotators. This paper presents a first look at this problem by using an existing Bayesian star formation inference package, \textsc{starpy}, to determine the quenching histories of a sample of quenching or quenched fast and slow rotators identified in the MaNGA sample, irrespective of visual morphology. We use broadband optical, $u-r$, and near-ultraviolet, $NUV-u$, colours from SDSS and GALEX to infer both the onset time and exponential rate of quenching for each galaxy. We aim to determine whether kinematically distinct galaxies have different quenching histories. 

This paper proceeds as follows. In Section~\ref{sec:datamethods} we describe our data sources and our Bayesian inference method for determining the quenching histories. We present our results in Section~\ref{sec:results} and discuss the implications of these results in Section~\ref{sec:discussion}. The zero points of all magnitudes are in the AB system. We adopt the WMAP Seven-Year Cosmology \citep{jarosik11} with $(\Omega_m , ~\Omega_\Lambda , ~h) = (0.26, 0.73, 0.71)$.

\section{Data and Methods}\label{sec:datamethods}

\subsection{SDSS \& GALEX Photometry}\label{sec:photom}

We use optical photometry from the Sloan Digital Sky Survey Data Release 7 (SDSS; \citealt{york00, abazajian09}). We use the Petrosian magnitude, {\tt petroMag}, values for the $u$ ($3543 \rm{\AA}$) and $r$ ($6231 \rm{\AA}$) wavebands provided by the SDSS DR7 pipeline \citep{stoughton02}. Further to this, we also required NUV ($2267 \rm{\AA}$) photometry from the GALEX survey \citep{martin05}. Observed fluxes are corrected for galactic extinction \citep{Oh11} by applying the \citet*{Cardelli89} law. We also adopt $k$-corrections to $z = 0.0$ and obtain absolute magnitudes from the NYU-VAGC \citep{blanton05, padmanabhan08, blanton07}.

\subsection{MaNGA Survey \& Data Reduction Pipeline}\label{sec:manga}

MaNGA is a multi-object IFU survey conducted with the $2.5~\rm{m}$ Sloan Foundation Telescope \citep{gunn06} at Apache Point Observatory (APO) as part of SDSS-IV \citep{blanton17}. By 2020 MaNGA will have acquired IFU spectroscopy for $\sim10000$ galaxies with $M_* > 10^9~\rm{M}_{\odot}$ and an approximately flat mass selection \citep{wake17}. The target selection is agnostic to morphology, colour and environment. 

MaNGA makes use of the Baryon Oscillation Spectroscopic Survey (BOSS) spectrograph \citep{smee13}. The BOSS spectrograph provides continuous coverage between $3600~\AA$ and $10300~\AA$ at a spectral resolution $R \sim 2000$ ($\sigma_{\rm{instrument}} \sim 77 \rm{km}~\rm{s}^{−1}$ for the majority of the wavelength range\footnote{Instrument resolution as a function of wavelength in shown in Figure 20 of \cite{yan16}}).

Complete spectral coverage to $1.5 R_e$, a galaxy's effective radius, is obtained for the majority of targets; a subset have coverage to $2.5 R_e$. See \cite{bundy15} for an overview of the MaNGA survey. For a further description of the instrumentation used by MaNGA see~\cite{drory15}. For a detailed description of the observing strategy see~\cite{law15} and for a description of the survey design see~\cite{yan16}. 

The raw data was processed by the MaNGA data reduction pipeline (DRP version 2.0.1), which is discussed in detail in \cite{law16}. The MaNGA DRP extracts, wavelength calibrates and flux calibrates all fibre spectra obtained in every exposure. The individual fibre spectra are then used to form a regular gridded datacube of $0.5''$ ‘spaxels’ and spectral channels. The spectra are logarithmically sampled with bin widths of $\log{\lambda} = 10^{-4}$. 

These datacubes are then analysed using the MaNGA data analysis pipeline (DAP version 2.0.2); the development of which is ongoing and will be described in detail in {\refcol Westfall et al. (\emph{in prep})}. {\minor Briefly, the spectral emission lines are masked, and the stellar continuum is modelled using the kinematic and stellar population fitting package p\textsc{pxf} \citep{cappellari04}. The stellar continuum model is then constructed using a thinned version of the MILES spectral library (wavelength range $3525 < \lambda~[\rm{\AA}] < 7500$). The model is broadened to match the stellar velocity dispersion of the galaxy in order to cleanly subtract the absorption lines from the spectrum. The residual emission lines are then modelled using Gaussian profiles, with 21 different lines fit in total.} The primary output from the DAP are therefore 2D ``maps" (i.e., images) of these measured properties, including flux, stellar and gas kinematics, spectral index measurements, and absorption- and emission-line properties. The effective radius of a galaxy and the ellipticity within it, $\epsilon_e$, are provided for MaNGA galaxies in the NASA Sloan Atlas; we use the values measured with elliptical Petrosian apertures in {\tt v1\_0\_1} of the catalogue provided in the SDSS Data Release 13 \citep{albareti16}. 

\subsection{Data sample}\label{sec:mangasample}

Our galaxy sample is drawn from the $2,777$ SDSS galaxies which make up the MaNGA DR14 data release \citep{dr14}. We cross-matched these galaxies with a radius of $3''$ to the GALEX survey in order to obtain NUV photometry (see Section~\ref{sec:photom}), resulting in $1,413$ galaxies.

In this study we wish to investigate the quenching histories of galaxies, therefore we sub-select those galaxies which are below the star forming sequence (SFS). Here we use the global average star formation rates (SFR) quoted in the MPA-JHU catalogue\footnote{\url{http://wwwmpa.mpa-garching.mpg.de/SDSS/DR7/}} \citep[][which are corrected for aperture bias]{kauffmann03, brinchmann04}. We do not use the MaNGA spectra to calculate SFRs; since the bundles only extend to $1.5~R_e$ we might miss star formation occurring in the outer regions of galaxies which would result in an underestimate of the global SFR of a galaxy. 

We select galaxies with a SFR more than $1\sigma$ below the SFS of \cite{peng10}. Since we wish to test whether slow rotators quench at rapid rates, consistent with major mergers, we wish to include those galaxies which have just left the SFS (rather than only selecting those that are fully quenched, for example, $3\sigma$ below the SFS).

\begin{figure}
\centering
\includegraphics[width=0.475\textwidth]{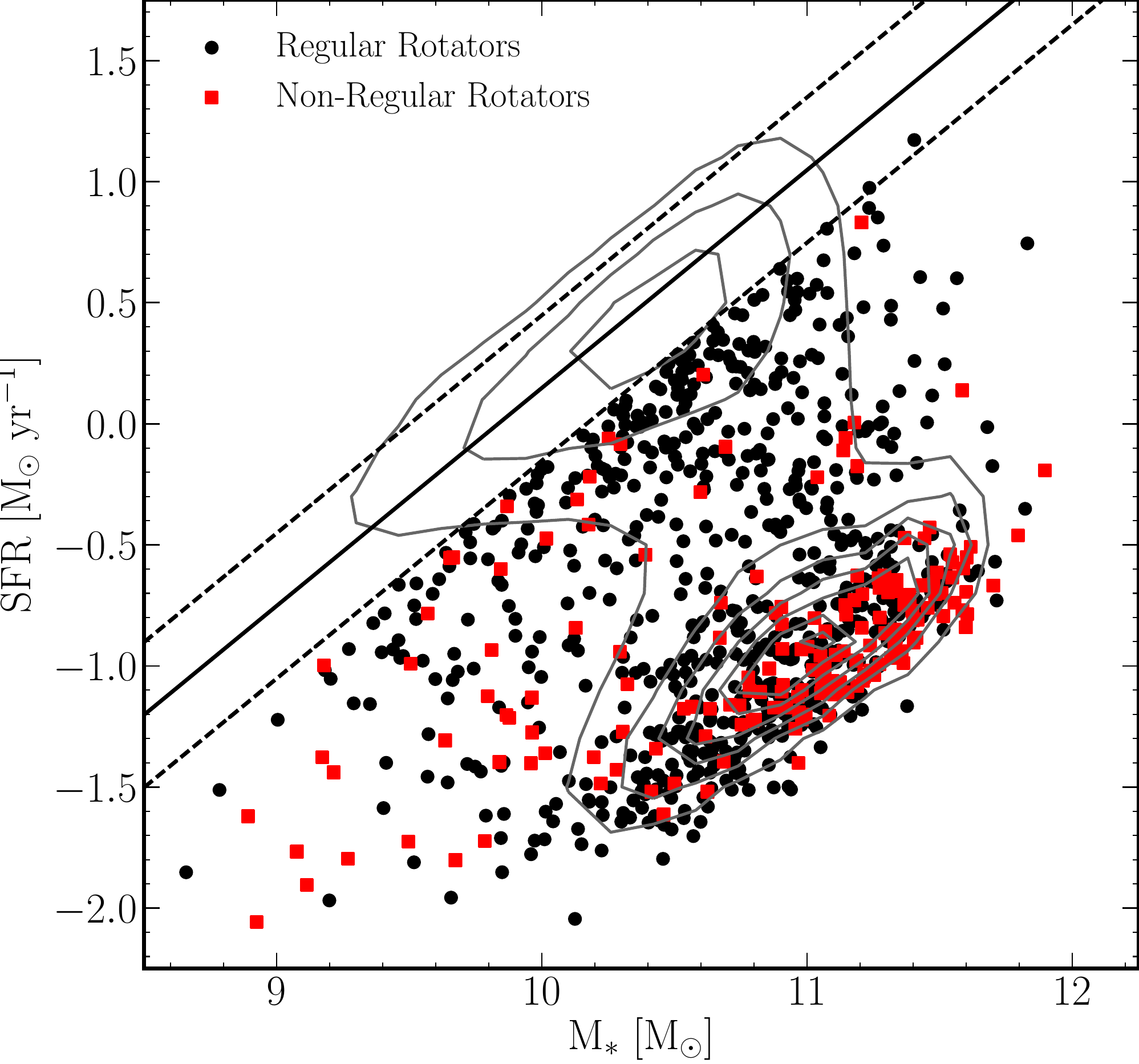}
\caption{Stellar mass against star formation rate for the \textsc{q-manga-galex} sample with regular (black circles) and non-regular (red squares) rotators identified using Equation~\ref{eq:fvs}. Shown also are the contours for the entire MPA-JHU sample (grey contours; i.e. SDSS DR7). The solid line shows the SFS as defined by \protect\cite{peng10} at the average redshift of the \textsc{q-manga-galex} sample, with $\pm 1 \sigma$ shown by the dashed lines. Note that the galaxies in the \textsc{q-manga-galex} sample are chosen to be more than $1\sigma$ below the SFS as defined at their observed redshift and stellar mass (see Section~\protect\ref{sec:mangasample}).}
\label{fig:masvsfr}
\end{figure}

This selection on SFR when applied to the \textsc{manga-galex} sample results in a sample of $826$ quenching or quenched galaxies, which we will refer to as the \textsc{q-manga-galex} sample. This sample is shown in Figure~\ref{fig:masvsfr}.


\subsection{Identifying Slow and Fast Rotators}\label{sec:fvs}

In order to classify the galaxies in the \textsc{q-manga-galex} sample as slow rotators or otherwise, we first calculate the specific stellar angular momentum as defined by \cite{emsellem07, emsellem11};
\begin{equation}
\lambda_{R_{e}} = \frac{\sum_{i=1}^{N} F_i\ R_i\ |V_i|}{\sum_{i=1}^{N} F_i\ R_i\ (V_i^2 + \sigma_i^2)^{1/2}},
\end{equation}	

where $F_i$ is the flux in the $i$th spaxel, $R_i$ the spaxel's distance from the galaxy centre (where $R_i < R_e$, the effective radius of a galaxy), $V_i$ the mean stellar velocity in that spaxel, $\sigma_i$ the stellar velocity dispersion in that spaxel and $N$ the total number of spaxels. In this work we use the \emph{Python} function provided in the MaNGA DAP to calculate $\lambda_{R_{e}}$ using the values of mean flux, radius, stellar velocity and stellar velocity dispersion (corrected for instrumental resolution effects) in each bin of the MaNGA data cubes binned with a signal-to-noise ratio of 10 using a Voronoi binning algorithm \citep{cappellari03}, as calculated by the MaNGA DAP (see Section~\ref{sec:manga}). Velocity dispersion measurements in each bin of a galaxy data cube were confirmed to be above the instrument resolution of $77~\rm{km}~\rm{s}^{-1}$.

We then classify galaxies in the \textsc{q-manga-galex} sample as non-regular rotators, or otherwise, using the definition from \cite{cappellari16}:
\begin{equation}\label{eq:fvs}
\lambda_{R_{e}} < 0.08 + \frac{\epsilon_e}{4} ~~~~~ \rm{with} ~~~~~ \epsilon_e < 0.4.
\end{equation}

\begin{figure*}
\centering
\includegraphics[width=\textwidth]{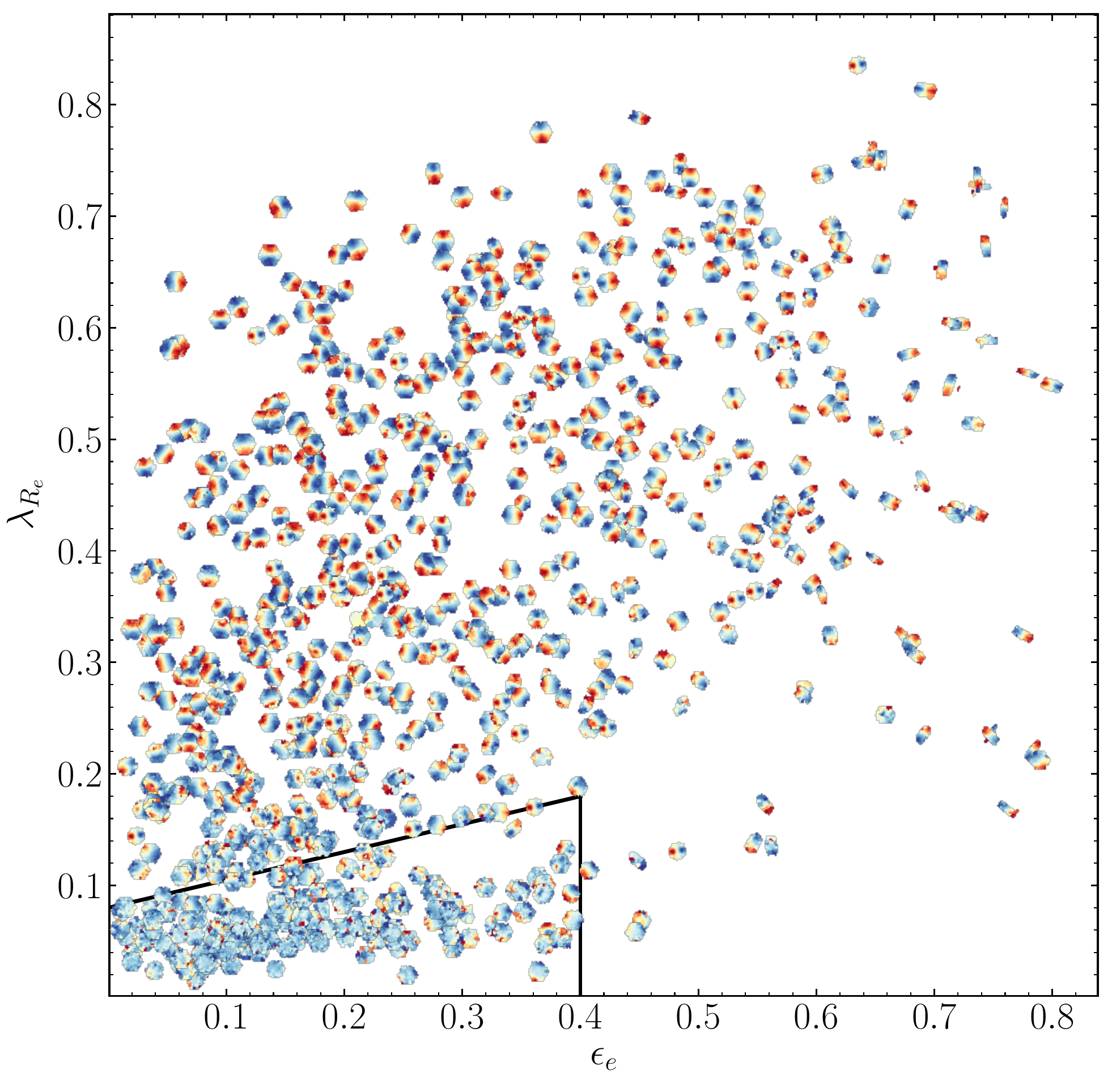}
\caption{Ellipticity versus stellar angular momentum for the regular and non-regular rotators of the \textsc{q-manga-galex} sample. Each point is shown by its stellar velocity map, each normalised to have a stellar velocity of $0~\rm{km}~\rm{s}^{-1}$ shown by the colour yellow. We show the separation between regular (i.e. fast) and non-regular rotators (i.e. slow rotators and objects with kinematically decoupled cores) from \protect\cite{cappellari16} with the solid black line.}
\label{fig:evsl}
\end{figure*}

Both slow rotators and kinematically disturbed galaxies will satisfy this inequality, hence why this selection results in a sample of non-regular rotators.  Using this definition reveals $168$ ($20\%$) non-regular rotators and $658$ ($80\%$) regular rotators in the \textsc{q-manga-galex} sample. Figure~\ref{fig:evsl} shows the velocity maps of these galaxies plotted at their values of $\lambda_{R_{e}}$ and $\epsilon_e$, along with the definition of a non-regular rotator from \cite{cappellari16} shown by the solid black line. Note the \textsc{q-manga-galex} sample is agnostic to visual morphology, so our sample of regular rotators will contain both rotationally supported early-types and late-type galaxies. 

The fraction of non-regular rotators found in the \textsc{q-manga-galex} sample ($20\%$) is slightly higher than that found by previous works \citep[$14-17\%$ of early-types in the $\rm{ATLAS}^{\rm{3D}}$ sample; ][]{emsellem11, stott16}. {\minor However, we must be wary with this comparison since the $\rm{ATLAS}^{\rm{3D}}$ sample is volume limited, whereas the MaNGA sample is selected to have a flat stellar mass distribution, prior to our selection on GALEX cross-matches and those galaxies below the SFS. Therefore although a direct comparison is not possible, we can at least determine if the fraction of non-regular rotators in the  \textsc{q-manga-galex} sample is a sensible figure given previous estimates.} Considering our sample is agnostic to visual morphology, {\minor we would  expect this selection effect to dominate resulting in a} smaller fraction of non-regular rotators than previous works which specifically derived the fraction of non-regular rotators in a sample of early-types only. However, many other studies have also shown that the non-regular rotator fraction increases with stellar mass \citep{cappellari13}, up to $\sim90\%$ at $10^{12}~\rm{M}_{\odot}$ \citep{veale17}. The median stellar mass of the \textsc{q-manga-galex} sample is $10^{10.8}~\rm{M}_{\odot}$, which is higher than the median stellar mass of the $\rm{ATLAS}^{\rm{3D}}$ sample at $10^{10.5}~\rm{M}_{\odot}$, likely accounting for this apparent discrepancy.

\begin{figure*}
\centering
\includegraphics[width=0.98\textwidth]{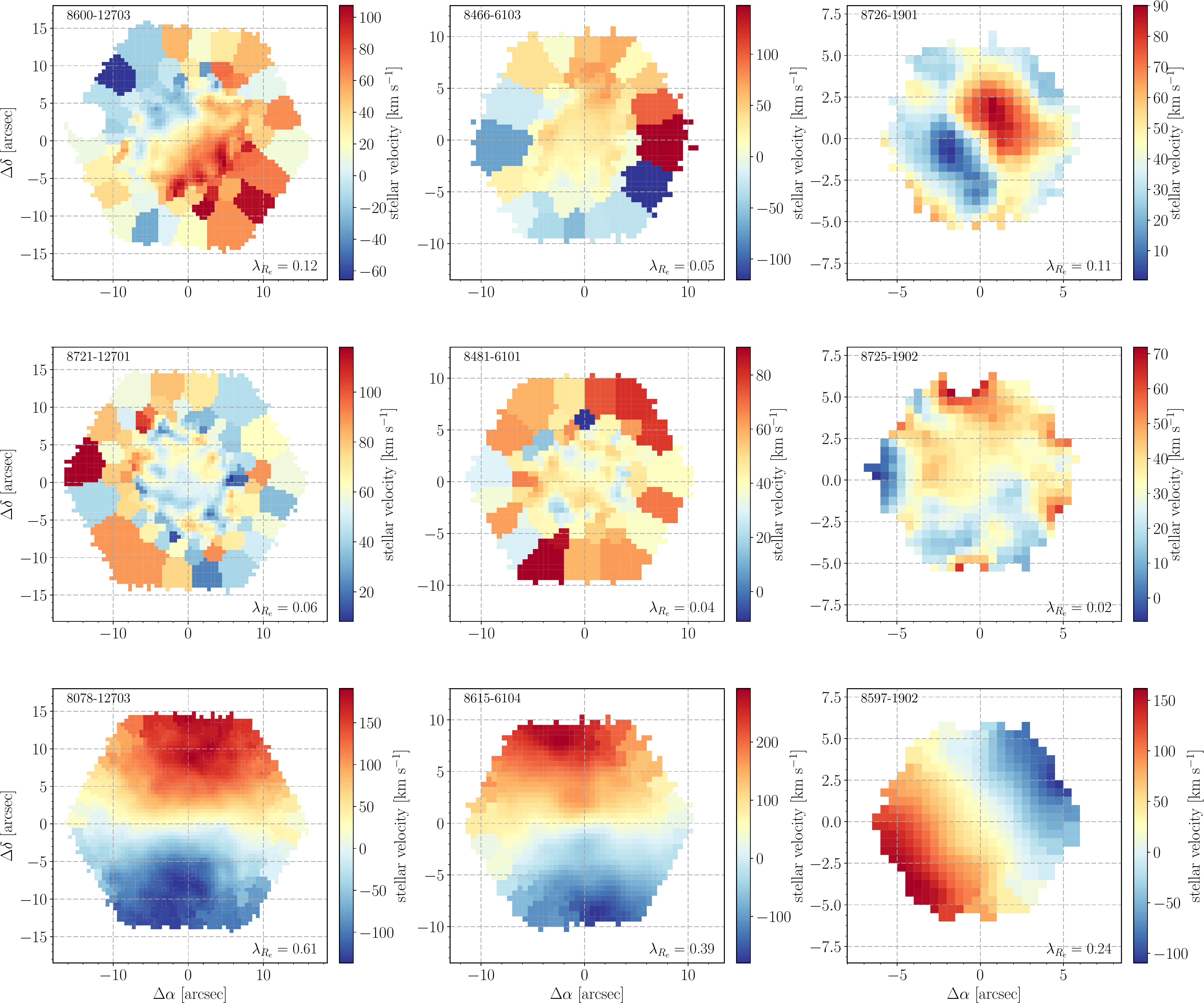}
\caption{Example stellar velocity maps, Voronoi binned with a signal-to-noise ratio of $10$, for three galaxies removed from the non-regular rotator \textsc{Q-MANGA-GALEX} sample because their kinematics show rotation (top row), three slow rotator galaxies without rotation (middle row), and three fast rotator galaxies with rotationally supported kinematics (bottom row). The MaNGA ID of each galaxy is shown in the top left of each panel and the measured $\lambda_{R_{e}}$ value in the bottom right. The number of spectral fibres in the MaNGA IFU bundle for each observation decreases from left to right.}
\label{fig:exvelmaps}
\end{figure*}  

In order to obtain a sample of slow rotators, one author (RJS) inspected the velocity maps of the $168$ non-regular rotators identified in the \textsc{q-manga-galex} sample to remove those galaxies which showed rotation in their kinematic map (i.e. counter rotation or decoupled cores). $71$ galaxies exhibiting rotation were identified, example velocity maps for which are shown in the top row of Figure~\ref{fig:exvelmaps}. This resulted in a sample of $97$ slow rotators, example velocity maps for which are shown in the middle row of Figure~\ref{fig:exvelmaps}.  

In order to control for the degeneracies between mass, metallicity and dust (all of which can redden a galaxy's optical colour and mimic the effects of quenching) we selected a sub-sample of fast rotators from those identified as regular rotators in the \textsc{q-manga-galex} sample. We matched to within $\pm~2.5~\%$ of the stellar mass of each slow rotator to give $97$ fast rotators, example velocity maps for which are shown in the bottom row of Figure~\ref{fig:exvelmaps}. We shall refer to this combined sample of $194$ fast and slow rotators as the \textsc{mm-q-manga-galex} sample. An Anderson-Darling \citep[AD;][]{anderson52} test reveals that the distribution of stellar masses of the fast rotators and slow rotators within this sample are statistically indistinguishable ($p=0.22$). Similarly their redshift distributions are also statistically indistinguishable ($p=0.19$).

\begin{figure}
\centering
\includegraphics[width=0.48\textwidth]{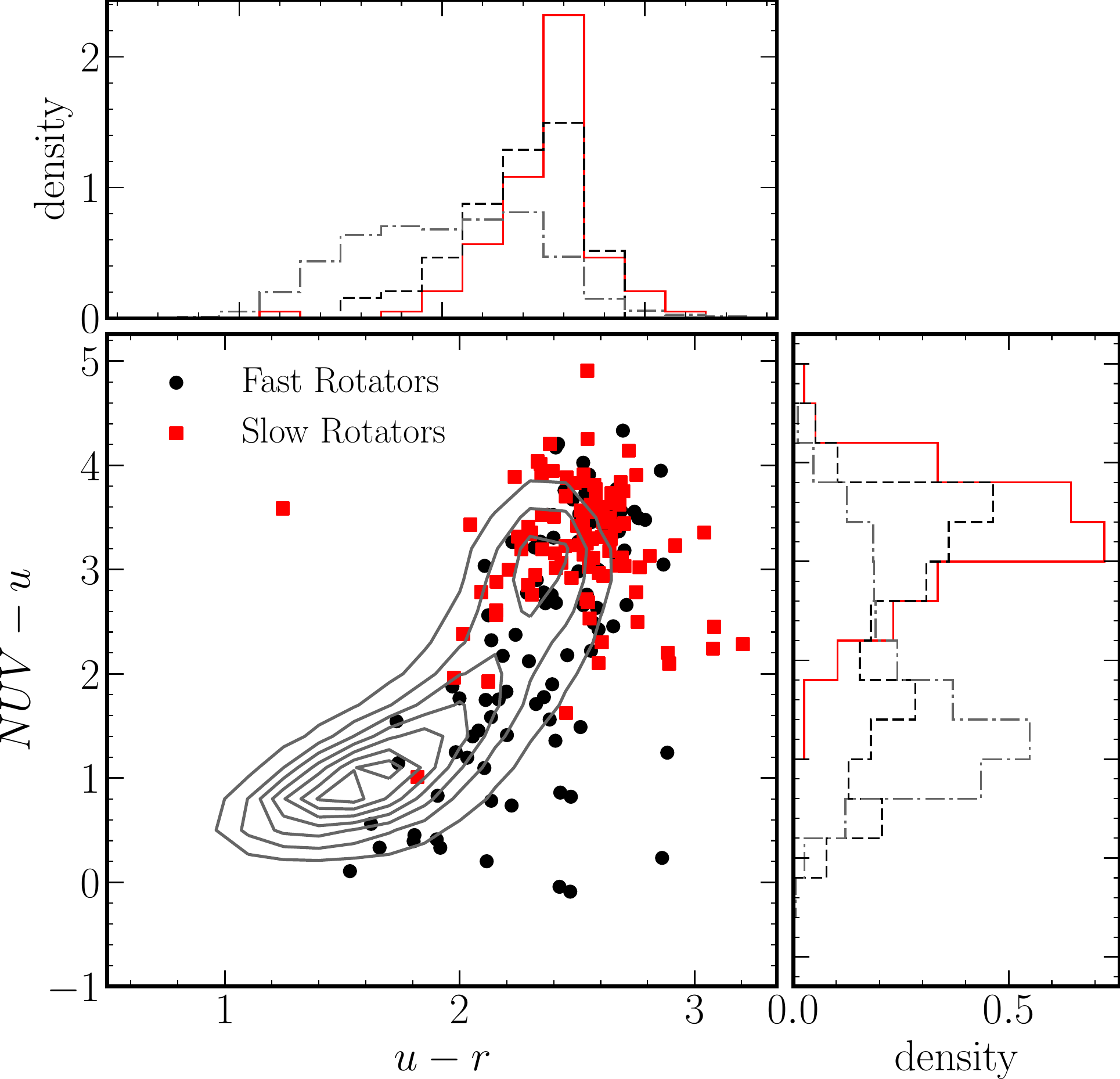}
\caption{Colour-colour diagram showing the optical $u-r$ and $NUV-u$ colours of the fast (black circles) and slow (red squares) rotators of the \textsc{mm-q-manga-galex} sample (main panel). Overlaid is the distribution of colours in a large SDSS-GALEX cross-matched sample from \protect\cite{smethurst15} for reference (grey contours). The top panel shows the normalised distribution of the $u-r$ colours of the fast (black dashed) and slow (red solid) rotators of the \textsc{mm-q-manga-galex} sample, along with the distribution for the \protect\cite{smethurst15} sample for reference (grey dot-dashed). Similarly the right panel shows the normalised distribution of the $NUV-u$ colours of the fast (black dashed) and slow (red solid) rotators of the \textsc{mm-q-manga-galex} sample, along with the distribution for the \protect\cite{smethurst15} sample for reference (grey dot-dashed).}
\label{fig:colcol}
\end{figure}

The optical and NUV colours from SDSS and GALEX (see Section~\ref{sec:photom}) for the \textsc{mm-q-manga-galex} sample are shown in Figure~\ref{fig:colcol}. Performing AD tests on the distributions of the colours of the slow and fast rotators in the \textsc{mm-q-manga-galex} sample reveals that both the $u-r$ ($AD= 5.9$, $p = 0.002$) and NUV-u ($AD= 19.1$, $p = 1\times10^{-5}$) colours of the two kinematic classifications are statistically distinguishable. These colours will be used to infer the SFHs of the \textsc{mm-q-manga-galex} sample (see Section~\ref{sec:starpy}).

\subsection{Environmental Densities}\label{sec:env}

We also consider the environmental densities of the fast and slow rotators by using estimates of the projected $5\rm{th}$ nearest neighbour density,  $\log\Sigma_5$, from \cite{bamford09}. An AD test reveals that the distribution of environment densities of the $72$ slow rotators and $80$ fast rotators of the \textsc{mm-q-manga-galex} sample with $\log\Sigma_5$ measurements from \cite{bamford09} are statistically indistinguishable ($p=0.28$). 

This is surprising since the current theory is that slow rotators are more likely to be the central galaxy of a group or cluster, whereas fast rotators are more likely to be satellite galaxies \citep{cappellari11, deugenio13, houghton13, scott14}. {\minor However, the MaNGA sample was chosen to be agnostic to galaxy environment, giving rise to a representative distribution of galaxy environments. Most galaxies in the sample will therefore reside in groups, a more common environment for a galaxy than the relatively rare environments of rich clusters \citep{carlberg04} or voids \citep{rieder13}. We must therefore probe the positions of the two samples within the group environment itself.} 

Cross-matching the \textsc{mm-q-manga-galex} sample with the \cite{yang09} SDSS group catalogue gives us group information for $94$ of the slow rotators and $96$ of the fast rotators. Similar fractions of these slow, $75/94$ ($80\%$), and fast rotators, $70/96$ ($73\%$), are classified as their brightest group galaxy (BGG). However, these fractions include those galaxies which are isolated in their halos {\minor (due to the theoretical definition of a BGG used in the \citealt{yang09} catalogue). These isolated galaxies could be the remains of a fossil group \citep{ponman94, jones00, jones03} or could be truly isolated, at the opposite end of the evolutionary spectrum which we are trying to probe. We must therefore remove these single galaxy `groups' in order to properly test whether the slow rotators are preferentially found at the centre of the groups in the \textsc{mm-q-manga-galex} sample.} 

Testing the distributions of the total group stellar mass for the fast and slow rotators we find they are statistically distinguishable (AD test $p=0.03$), with slow rotators residing in more massive groups. If we then consider only those galaxies in groups with a total stellar mass greater than $10^{11}±\rm{M}_{\odot}$ (under the simplifying assumption that this will remove the majority of single galaxy `groups') we find the fraction of slow rotators classified as a BGG is $44/61$ ($72\%$), whereas for fast rotators this drops to $30/52$ ($58\%$), a statistically distinguishable difference ($p=0.04$). Therefore, although the projected local environment densities of the two kinematic classes of galaxies are statistically indistinguishable, their positions within that given environment density do differ, as expected. 

Given the above statistical tests, the only differences between the fast and slow rotators of the \textsc{mm-q-manga-galex} sample is their kinematics, their colours and their position within their group halo.


\subsection{Star Formation History Inference}\label{sec:starpy}

\textsc{starpy}\footnote{Publicly available: \url{http://github.com/zooniverse/starpy}} is a \textsc{python} code which allows the inference of the exponentially declining star formation history (SFH) of a single galaxy using  Bayesian Markov Chain Monte Carlo techniques \citep{emcee13}\footnote{\url{http://dan.iel.fm/emcee/}}. The code uses the solar metallicity stellar population models of \cite[][hereafter BC03]{BC03}, assumes a Chabrier IMF \citep{chabrier03} and requires the input of the observed $u-r$ and $NUV-u$ colours and redshift. No attempt is made to model for intrinsic dust. 

The SFH is described by an exponentially declining SFR described by two parameters; the time at the onset of quenching, $t_q~\rm{[Gyr]}$, and the exponential rate at which quenching occurs, $\tau~\rm{[Gyr]}$. Under the simplifying assumption that all galaxies formed at $t=0$ $\rm{ Gyr}$ with an initial burst of star formation, the SFH can be described as:
\begin{equation}\label{sfh}
SFR =
\begin{cases}
i_{sfr}(t_q) & \text{if } t < t_q \\
i_{sfr}(t_q) \times exp{\left( \frac{-(t-t_{q})}{\tau}\right)} & \text{if } t > t_q 
\end{cases}
\end{equation}
where $i_{sfr}$ is an initial constant star formation rate dependent on $t_q$ \citep{schawinski14, smethurst15}. The simplifying assumption that all galaxies formed at $t~=~0$~$\rm{Gyr}$ means that the age of each galaxy, $t_\mathrm{age}$, corresponds to the age of the Universe at its observed redshift, $t_\mathrm{obs}$. A smaller $\tau$ value corresponds to a rapid quench, whereas a larger $\tau$ value corresponds to a slower quench. A galaxy undergoing a slow quench is not necessarily quiescent by the time of observation. This SFH model has previously been shown to appropriately characterise quenching galaxies \citep{Weiner06, Martin07, Noeske07,schawinski14}. 

The probabilistic fitting methods to these star formation histories for an observed galaxy are described in full detail in Section 3.2 of \cite{smethurst15}, wherein the \textsc{starpy} code was used to characterise the morphologically dependence of the SFHs of $\sim126,000$ galaxies. Similarly, in \cite{smethurst16}, \textsc{starpy} was used to show the prevalence of rapid, recent quenching within a population of AGN host galaxies and in \cite{smethurst17} to investigate the quenching histories of group galaxies.  

\begin{figure*}
\centering
\includegraphics[width=\textwidth]{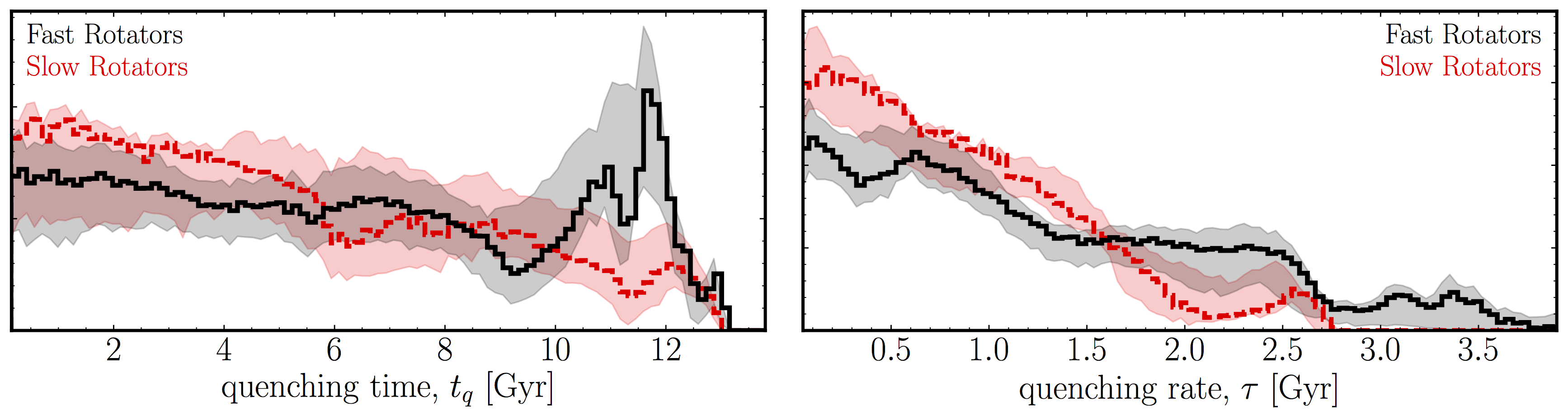}
\caption{Population densities for the time, $t_q$ (left) and exponential rate, $\tau$ (right) that quenching occurs in the \textsc{mm-q-manga-galex} sample for the fast (black, solid) and slow (red, dashed) rotators. A high value of $t_q$ corresponds to a recent quench, and a high value of $\tau$ corresponds to a slow quench. Shaded regions show the uncertainties on the distributions from bootstrapping. 
}
\label{fig:popfrvsr}
\end{figure*}

Briefly, we assume a flat prior on all the model parameters and model the difference between the observed and predicted $u-r$ and $NUV-u$ colours as independent realisations of a double Gaussian likelihood function (Equation 2 in \citealt{smethurst15}). An example posterior probability distribution output by \textsc{starpy} is shown for a single galaxy in Figure 5 of \cite{smethurst15}, wherein the degeneracies of the SFH model between recent, rapid quenching and earlier, slower quenching can be seen.

To study the SFH across a sample of many galaxies, these individual posterior probability distributions are stacked in $[t_q, \tau]$ space to give one distribution across each quenching parameter for the sample. This is no longer inference but merely a method to visualise the results for a population of galaxies (see appendix section C in \citealt{smethurst16} for a discussion on alternative methods which may be used to determine the parent population SFH). These distributions will be referred to as the population SFH densities.

\section{Results}\label{sec:results}

We determine the population SFH densities for both the fast and slow rotators of the \textsc{mm-q-manga-galex} sample. This is shown in Figure~\ref{fig:popfrvsr} for both the onset time (left panel) and exponential rate (right panel) of quenching for the fast (black solid line) and slow (red dashed line) rotators. Uncertainties on the population densities (shown by the shaded regions) are determined from the maximum and minimum values spanned by $N = 1000$ bootstrap iterations, each sampling $90\%$ of either the fast (black shaded region) or slow (red shaded region) rotators. 

To statistically test the significance of our results, we estimate the `best fit' $[t_q, \tau]$ values for each galaxy with the median value of an individual galaxy's posterior probability distribution from \textsc{starpy} (i.e. the 50th percentile position of the MCMC chain). We test the distribution of these values of the fast and slow rotators in the \textsc{mm-q-manga-galex} sample with AD-tests. Firstly, an AD-test on the distributions of $t_q$ values in the fast and slow rotator samples, revealed that we cannot reject the null hypothesis that the fast and slow rotators quench at the same time ($AD= 0.65$, $p = 0.69$). Finally, an AD-test on the distributions of $\tau$ values, revealed that we can reject the null hypothesis that the fast and slow rotators quench at the same rate ($AD= 6.3$, $p = 0.001$). This is a $3.2\sigma$ result which suggests that slow rotators quench faster than fast rotators of the same mass.

\section{Discussion}\label{sec:discussion}

The results presented in Section~\ref{sec:results} suggest that fast and slow rotators are indeed separate populations quenched, and therefore formed, by different mechanisms. However, these quenching mechanisms occur at statistically indistinguishable onset times for fast and slow rotators. \cite{khochfar11} find in their simulations that the last major merger interaction for slow rotators was at $z \gtrsim 1.5$ (i.e. $t_q \lesssim 4.5~\rm{Gyr}$). However, \cite{penoyre17}  find in the Illustris simulation that slow rotators only form after $z < 1$ (i.e. $t_q \gtrsim 6~\rm{Gyr}$). We note that \textsc{starpy} is not very sensitive to the time of quenching, particularly at early times ($t_q \lesssim 6~\rm{Gyr}$ when $z \gtrsim 1$), due to the degeneracies between the optical and NUV colours currently used to infer the quenching parameters. Therefore, we cannot currently conclude which scenario our results favour. Future work altering our inference code to take spatial spectral information provided by MaNGA may help us to address this issue by breaking the degeneracies inherent in the photometric colours.

However, \textsc{starpy} in its current form is sensitive to the rate of quenching in a galaxy. In the right panel of Figure~\ref{fig:popfrvsr} we see that there is a wide range of quenching rates occurring within the fast rotator sample. Previous works using \textsc{starpy} have shown how the intermediate quenching rates ($1 \lesssim \tau~[\rm{Gyr}]\lesssim 2$) prevalent in the distribution of the fast rotator sample can be attributed to environmental processes such as harassment and galaxy interactions \citep{smethurst17}, or minor mergers \citep{smethurst15}. This is unsurprising given that the fast rotators are less likely to be the brightest group galaxy than the slow rotators of the \textsc{mm-q-manga-galex} sample, as discussed in Section~\ref{sec:env}.

In particular we find evidence for galaxies in the fast rotator sample to quench at slow rates ($\tau \geq 2~\rm{Gyr}$). Since the \textsc{q-manga-galex} sample is agnostic to visual morphology, it will contain fast rotators which are disc dominated (i.e. late-type galaxies). This preference for slow quenching rates is therefore likely to be caused by the effects of secular evolution through gas accretion and morphological quenching, slowly moving these disc galaxies off the SFS to produce the red spiral population of \cite{masters12a}. Using the morphological classifications of Galaxy Zoo 2 \citep[GZ2][]{lintott11, GZ2} we find that $20/97$ ($21\%$) of the fast rotators of the \textsc{mm-q-manga-galex} sample are disc dominated with a disc or featured debiased vote fraction, $p_d \geq 0.8$ (i.e. $80\%$ of classifiers marked the galaxy as having either a disc or features). This is consistent with the fact that $23\pm^{2}_{11}\%$ of the fast rotator quenching rate population density (black line in the right panel of Figure~\ref{fig:popfrvsr}) is found at quenching rates $\tau > 2~\rm{Gyr}$. 

Conversely only 1 of the slow rotators was classified as having a disc or features by GZ2\footnote{Upon visual inspection this galaxy has a large disc with spiral structure lying outside of the MaNGA fibre bundle at $>1.5~\rm{R_e}$}. It is not surprising therefore, that there is much less preference for slow quenching rates, with $\tau \geq 2~\rm{Gyr}$, for slow rotators than fast rotators in the right panel of Figure~\ref{fig:popfrvsr}. However, \cite{smethurst15} found for galaxies in the red sequence visually classified as `smooth' in GZ2 (i.e. quenching or quenched early-types) that a significant fraction, $26.1\%$, of the quenching rate population density was found at these slow quenching rates (see left panel of their Figure 8). However, a sample of  visually classified `smooth' galaxies in GZ2 may include both fast and slow rotators. It is only in this work that we have been able to investigate the difference in the SFHs of galaxies which are rotationally supported from those which are not, revealing that the stellar kinematics are driving the morphologically dependant star formation histories seen in \cite{smethurst15}.

The slow rotators in the \textsc{mm-q-manga-galex} sample instead show a preference for rapid quenching rates ($\tau\lesssim1~\rm{Gyr}$) in the right panel of Figure~\ref{fig:popfrvsr}. Assuming that major mergers are the only mechanism able to destroy rotation in a galaxy, this result supports the theory that these galaxies are formed by major mergers which, along with destroying the disc of a galaxy, are thought to cause quenching at such rapid rates ~\citep{springel05b, bell06, lotz08b,lotz11}. Surprisingly, we also find evidence that some of the fast rotators are quenching at these same rapid rates ($\tau \lesssim 1~\rm{Gyr}$) in the right panel of Figure~\ref{fig:popfrvsr}. This suggests that in a fraction of fast rotators a dynamically fast process, such as a major merger, may be the cause of quenching. 

Simulations have recently shown that although major mergers (2:1 or 1:1 mergers) can cause rapid quenching of a galaxy, they do not necessarily destroy the disc dominated nature of a galaxy ~\citep{pontzen16, sparre16} and can actually form a fast rotator remnant \citep{bois11}. This is thought to mainly occur in gas rich major mergers \citep{bois11} and is likely the explanation for the presence of rapid rates in the fast rotator sample seen in the right panel of Figure~\ref{fig:popfrvsr}. We therefore predict that the fast rotators in the \textsc{mm-q-manga-galex} sample will be more gas rich than the slow rotators they are stellar mass matched to. We will be able to test this hypothesis with currently ongoing follow-up observations using the Green Bank Telescope (GBT16A-095 and GBT17A-012; {\refcol Masters et al. \emph{in prep.}}) which will obtain HI profiles for galaxies in the MaNGA target sample. With these observations we will be able to determine whether gas mass has an impact on the formation mechanisms of these kinematically distinct galaxies. 


\section{Conclusions}

We have investigated the star formation histories of quenching or quenched fast and slow rotators identified in the MaNGA galaxy sample, irrespective of their visual morphology. We used the $u-r$ and $NUV-u$ colours with an existing piece of inference software, \textsc{starpy}, to determine the onset time and exponential rate of quenching in each of these galaxies. 

An Anderson-Darling test revealed that the distribution of the inferred quenching rates of fast and slow rotators are statistically distinguishable ($p=0.001$, $3.2\sigma$). We find that rapid quenching rates ($\tau \lesssim 1~\rm{Gyr}$) are dominant for slow rotators, supporting the theory that slow rotators form in dynamically fast processes, such as major mergers \citep{bois10, duc11, naab14}. Conversely, we find that fast rotators quench at a wide range of rates, consistent with dynamically slow processes such as secular evolution, minor mergers, gas accretion and environmentally driven mechanisms. However we also find evidence that some of the fast rotators are quenching at the same rapid rates dominant across the slow rotator sample.

This finding of rapid quenching rates occurring for both slow rotators and a subset of the fast rotators suggests that although their kinematics are different in nature, both classes of galaxy may be able to quench, and therefore form, via major mergers. This result combined with the findings of recent simulations showing disc survival in gas-rich major mergers ~\citep{bois11, pontzen16, sparre16}, suggests that the total gas mass fraction within a pair of merging galaxies, is what will ultimately decide the kinematic fate of a galaxy. 
	
\section*{Acknowledgements}

RJS gratefully acknowledges research funding from the Ogden Trust. AW acknowledges support of a Leverhulme Trust Early Career Fellowship.

Based on observations made with the NASA Galaxy Evolution Explorer.  GALEX is operated for NASA by the California Institute of Technology under NASA contract NAS5-98034.

Funding for the Sloan Digital Sky Survey IV has been provided by the Alfred P. Sloan Foundation, the U.S. Department of Energy Office of Science, and the Participating Institutions. SDSS acknowledges support and resources from the Center for High-Performance Computing at the University of Utah. The SDSS web site is \url{www.sdss.org}.

SDSS is managed by the Astrophysical Research Consortium for the Participating Institutions of the SDSS Collaboration including the Brazilian Participation Group, the Carnegie Institution for Science, Carnegie Mellon University, the Chilean Participation Group, the French Participation Group, Harvard-Smithsonian Center for Astrophysics, Instituto de Astrof\'isica de Canarias, The Johns Hopkins University, Kavli Institute for the Physics and Mathematics of the Universe (IPMU) / University of Tokyo, Lawrence Berkeley National Laboratory, Leibniz Institut für Astrophysik Potsdam (AIP), Max-Planck-Institut f\"ur Astronomie (MPIA Heidelberg), Max-Planck-Institut für Astrophysik (MPA Garching), Max-Planck-Institut f\"ur Extraterrestrische Physik (MPE), National Astronomical Observatories of China, New Mexico State University, New York University, University of Notre Dame, Observat\'orio Nacional / MCTI, The Ohio State University, Pennsylvania State University, Shanghai Astronomical Observatory, United Kingdom Participation Group, Universidad Nacional Aut\'onoma de M\'exico, University of Arizona, University of Colorado Boulder, University of Oxford, University of Portsmouth, University of Utah, University of Virginia, University of Washington, University of Wisconsin, Vanderbilt University and Yale University.

\bibliographystyle{mn2e}
\bibliography{refs}  




\end{document}